\documentclass[printer]{aa}
\usepackage{txfonts}
\usepackage{graphicx}

\begin{document}
\title{The angular momentum of condensations within elephant trunks}
\author{V. Lora \inst{1}
\and A. C. Raga \inst{2}
\and A. Esquivel\inst{2}}

\institute{Instituto de Astronom\'ia, Universidad Nacional Aut\'onoma de M\'exico, Ap. 70-468,
04510, D.F M\'exico
\and Instituto de Ciencias Nucleares, Universidad Nacional Aut\'onoma de M\'exico, Ap. 70-543, 
04510, D.F M\'exico}
\date{Received date /
       Accepted date}
\abstract 
{}
{The radiation from newly born stars photoevaporates their parental 
neutral cloud, leading to the formation of dense clumps that will eventually form stars.}
{We present 3D simulations of the interaction of a neutral cloud with an external 
ionising radiation field, and compute the angular momenta of these collapsing clumps.}
{The angular momenta of these collapsing clumps show that they
have preferential 
orient mostly perpendicular to the
direction of the incident ionising photon field. Therefore, the axes
of the jet systems that will be eventually ejected
(from the $star+accretion$ $disk$ systems that will form) will
be oriented approximately perpendicular to the direction to the
photoionising source. }
{}

\maketitle

\titlerunning{Angular momentum in elephant trunks}
\authorrunning{V. Lora et al. }

\section{Introduction}
The radiation from newly born stars photoionises and erodes the parental
cloud, producing structures such as the so-called elephant trunks.
At the head of an elephant trunk, the interaction of the shock (driven
by the photoevaporation process) with previously existing density
perturbations leads to the formation of dense clumps. Some of these
clumps might have enough mass to be self-gravitating, and will eventually
form young stars that eject bipolar outflows. We describe
observed examples of this kind of configuration.

Bally \& Reipurth ($2003$) discovered HH objects in the molecular cloud
associated with the Pelican Nebula, including \object{HH~555}. This outflow
emerges from the tip of a long elephant trunk, providing direct evidence
of ongoing star formation in this region.
The outflow axis of \object{HH~555} is approximately perpendicular to the
elephant trunk, which is aligned with the direction to the
photoionising source.

Another example of this kind of configuration is the \object{HH~666} outflow in
the Carina nebula. \object{HH~666} also emerges from close to the tip
of an elephant trunk, and its axis is almost perpendicular to the
direction towards $\eta$ Carinae (Smith, Bally \& Brooks 2004).
An HST image of this region (Bally, Reipurth \& Davis 2007),
shows a second jet emerging from a nearby elephant trunk, with
a direction almost parallel to the \object{HH~666} outflow.

A final example is provided by \object{HH~333}. This jet emerges from
the tip of an elephant trunk within the complex Trifid nebula
(\object{M20}). It is a single-sided jet with measured radial
velocities (Rosado et al. 1999) and proper motions
(Yusef-Zadeh et al. 2005) 
that indicate it has the kinematical properties of a
standard HH jet. Again \object{HH~333} has an outflow direction
approximately perpendicular to the direction to
the ionising source.

Reach et al. (2009) presented observations of the
elephant trunk of globule \object{IC~1396A}. They detected outflow activity
from a number of young stars in the region. However,
it is impossible to determine the
outflow axes from these observations.

Even though the number of four outflows (two
in the \object{HH~666} region, see above) observed to be emerging
from tips of elephant trunks is quite small, the
alignment approximately perpendicular
to the direction to the ionising photon source might
be indicative of a systematic alignment.
This alignment implies that the angular
momenta of the low mass star+disk systems producing outflows
from stellar sources in the tip of elephant trunks
are more or less perpendicular to the direction of the ionising
photon field (produced by the massive stars giving rise to
the photoionised nebulae and elephant trunk structures). These 
angular momenta presumably preserve the direction of the rotation
axes of the dense clumps that collapsed to form the outflow sources.

In the present paper, we explore the interaction between an
ionising photon field and an environment with density perturbations.
This interaction produces elongated structures reminiscent of
elephant trunks, with dense, embedded clumps.
In particular, we focus on 
whether or not these dense clumps have angular momenta preferentially
oriented perpendicular to the direction towards the photoionising source.

Mellema et al. (2006) carried out 3D, radiation gasdynamic simulations
of an H~II region expanding in an ISM with power-law density perturbations.
They find that this configuration naturally leads to the formation of
dense, radially elongated structures, which resemble elephant trunks. Also,
Gahm et al. (2006) study the role of magnetic fields in the formation
of elephant trunks. Finally, Gritschneder et al. (2009) carried
out a simulation of an initially plane ionising front travelling
into a structured medium.

Our work emulates the approach of Mellema et al. (2006)
and Gritschneder et al. (2009). We focus
on a small region of the edge of an expanding H~II region, and carry
out a 3D radiation gasdynamic simulation (including the self-gravity
of the gas) of the formation of a dense,
neutral structure. We then identify high density clumps within this
``elephant trunk'', and compute their angular momenta. Finally, we
study the mass distribution of the clumps, and the distributions
of the orientation and magnitude of their angular momenta.

The paper is organized as follows.
In Sect. $2$, we describe the gasdynamic code and the parameters
used for the numerical simulation. The results from the simulation
and the clump statistics are presented in Sect. $3$. Finally,
our results are summarised in Sect. 4.

\section{Code \& settings}
\subsection{Code}
We carried out a 3D simulation with a code that solves the
3D gasdynamic equations, the Poisson equation for the gravitational
field, and a rate equation for neutral hydrogen,
including the transfer of ionising photons at the Lyman limit.
The gas is initially atomic, and the models do not consider the
photodissociation of molecular material because of the presence
of a FUV radiation field.
This code was described by Raga et al. (2008).

We modified the code of Raga et al. (2008) to include
the ``two temperature'' equation of state described by 
Esquivel \& Raga (2007, hereafter E07). This equation of state
assigns temperatures between 10~K (for neutral gas) and $10^4$~K
(for gas with fully ionised H) with a linear dependence on the
H ionisation fraction. Therefore, instead of solving an energy equation
with the appropriate heating and cooling terms (see Raga et al. 2008),
we replace it with this two-temperature equation of state.
We also included the self-gravity of the gas. We use a
successive over relaxation (SOR) method to solve the Poisson
equation for the gravitational potential, and then include the
gravitational force in the momentum and energy equations.
We do not include a treatment of the diffuse, ionising photon field.

\subsection{Settings}
The computational domain has a size of $(3.0,1.5,1.5)\times 10^{18}$~cm
(along the $x$-, $y$- and $z$-axes, respectively),
which is resolved with a uniform grid of $256\times 128\times 128$ grid
points. We impose transmision boundaries in the $x$-direction and periodic
boundaries along the $y$ and $z$-directions. The periodic conditions
are imposed in the gasdynamic equations, in Poisson's equation
(for the gravitational field), and in the radiative transfer equations.

We start with an inhomogeneous density structure with a 
power-law power-spectrum index of $-11/3$ (i.e. $P[k]\propto k^{-11/3}$, 
where $k$ is the wave-number), as described in Esquivel et al. (2003).  
The initial density structure does not have any motion.
To simulate the edge of an H{\sc II} region, the computational 
domain is divided into two portions with a dividing line  
at $x=4\times 10^\mathrm{17}$~cm from the left edge of the domain.  
The portion to the left is filled with an ionised medium (with a 
temperature of $10^4~\mathrm{K}$, and the portion to the right 
is filled with a neutral medium (with a temperature of $10~\mathrm{K}$). 
The average density in the neutral medium is a factor of $100$ higher 
than the one in the ionised medium, and the transition between the two 
(also in terms of temperature and ionisation fraction) follows a $\tanh$ 
profile with a width of $\sim$ 10 pixels. The resulting neutral
structure has a mass of $228~\mathrm{M}_{\odot}$.

To calculate the gravitational field, we only consider the
gravitational force resulting from the density perturbations.
In other words, we subtract a density $\rho_0=3.51\times10^{-24}$~g~cm$^{-3}$
(corresponding to the lower density regions in the initial
distribution of neutral material)
from the density used in Poisson's equation. In this way, we
avoid a generalized collapse of the dense slab structure that
fills the computational domain. We also run a simulation
in which the gravitational force was ``turned off'' 
to illustrate the effect of the self-gravity of the gas.
 
A plane-parallel ionising photon field $F_0=8.8\times10^{10}$cm$^{2}$s$^{-1}$
is incident on the computational
domain along the $x$-axis. This photon flux corresponds to a star
with an ionising photon rate $S_*=10^{48}$~s$^{-1}$, located at a distance
$D=9.5\times10^{17}$~cm from the edge of the computational domain
in the $-x$ direction.


\begin{figure}
   \centering
   \includegraphics[width=8cm]{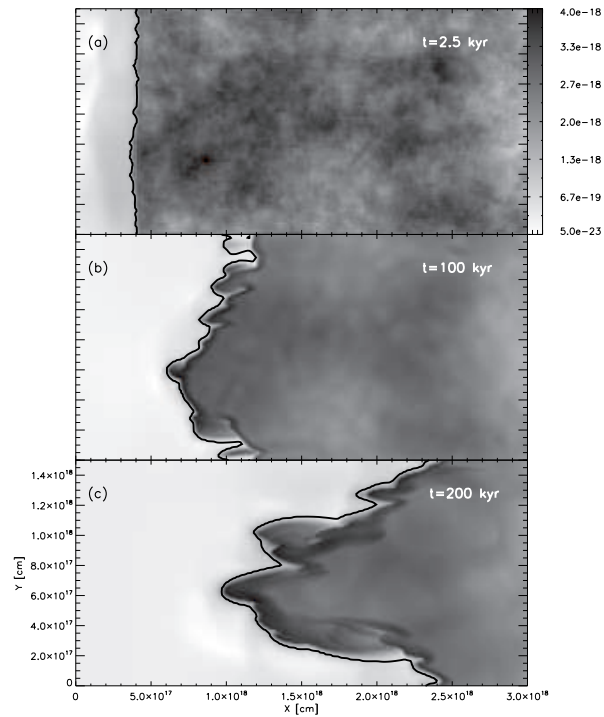}
\caption{Time evolution of the $xy$ mid-plane density stratification without
self-gravity. The three frames are labeled with the corresponding 
elapsed times. The density stratifications are shown with the 
logarithmic greyscale given (in g~cm$^{-3}$) by the top right bar. 
In the three frames we also show the contour corresponding to an H 
ionisation fraction of 50\%, which indicates the position of the 
ionisation front. The $x$ and $y$-axes are labeled in cm.}
\end{figure}
\begin{figure}
   \centering
   \includegraphics[width=8cm]{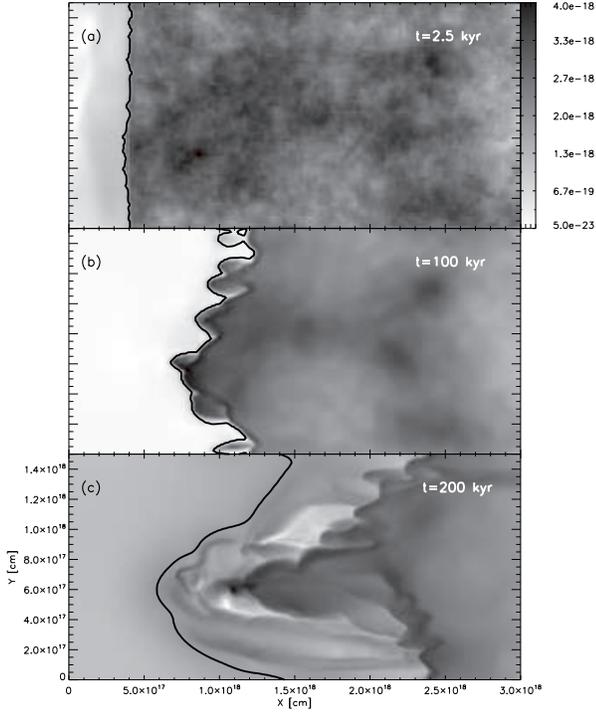}
\caption{Same as Fig. 1, but for the simulation that
includes the self-gravity of the gas.}
\end{figure}

\begin{figure}
   \centering
   \includegraphics[width=6cm]{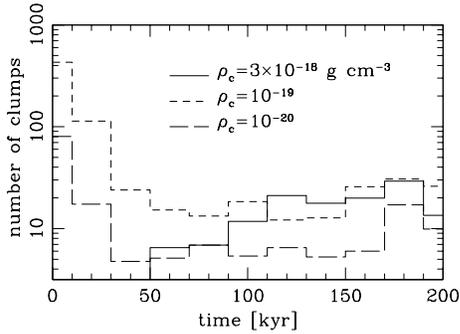}
\caption{Number of neutral clumps as a function of time, obtained
   for three different density cuttoffs. The results correspond
to the simulation which includes self-gravity (see Fig. 2).}
\end{figure}

\begin{figure}
  \centering
  \includegraphics[width=5cm]{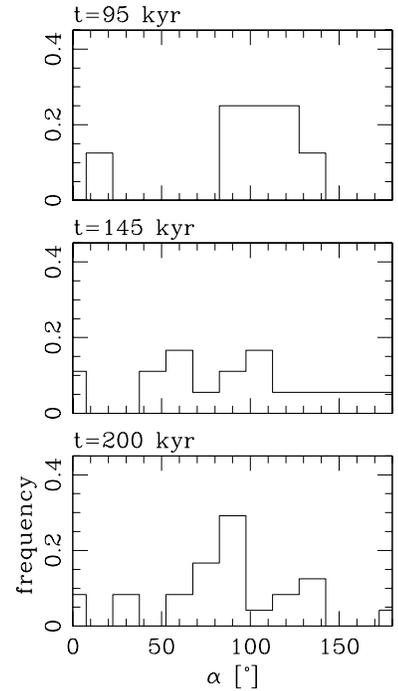}
  \caption{Fraction of the clumps (obtained for a cutoff density  
$\rho_c~=3\times10^{-18}$g~cm$^{-3}$) with different orientations $\alpha$
(between the $x$-axis and the $xy$-projections
of the angular momenta of the clumps). The panels
are labeled with the elapsed time corresponding to the time-frames
from which the three angular distributions were obtained. The results correspond
to the simulation that includes self-gravity (see Fig. 2).}
\end{figure}

\begin{figure}
   \centering
   \includegraphics[width=6cm]{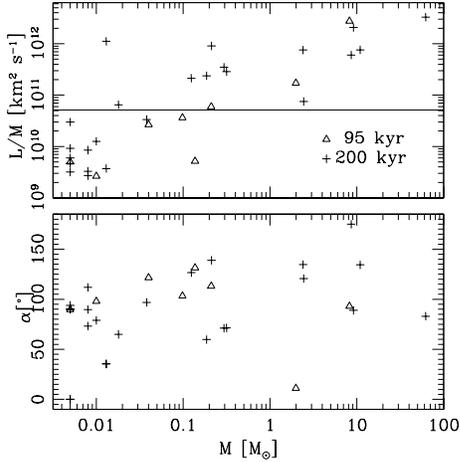}
\caption{The top panel shows the specific angular momentum $(L/M)$ of each clump
as a function of the clump's mass. The black line shows the angular momentum
associated with the outer orbit of an accretion disc radius of $r_{D}=50$~AU
around a $M=2M_{\odot}$ star ($L/M\sim5\times10^{10}$~km$^{2}s^{-1}$,
see the text).
The bottom panel shows the orientation $\alpha$ (between the $x$-axis
and the $xy$-projection of the angular momentum) for each clump, as a 
function of its mass. The triangles show the clumps found for an
elapsed time of $95$~kyr and the crosses for an elapsed time of
$195$~kyr. The clumps were obtained using a
$\rho_c~=3\times 10^{-18}$g~cm$^{-3}$ density cutoff, using the
results from the simulation of Fig. 2.}
\end{figure}

\begin{figure}
   \centering
   \includegraphics[width=5.5cm]{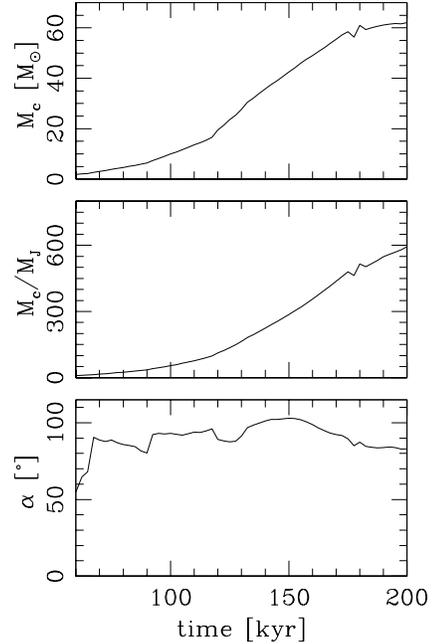}
\caption{Time evolution in the most massive, neutral clump obtained
from the simulation with self-gravity (see Fig. 2).
The top panel shows the mass, and the central panel shows
the ratio of the clump mass to the Jeans mass as a function of time.
The bottom panel shows the time evolution in 
orientation $\alpha$ of the angular momentum of this clump (where
$\alpha$ is the angle between the $x$-axis and the $xy$-plane projection of the
angular momentum).}
\end{figure}

\begin{figure}
   \centering
   \includegraphics[width=5.5cm]{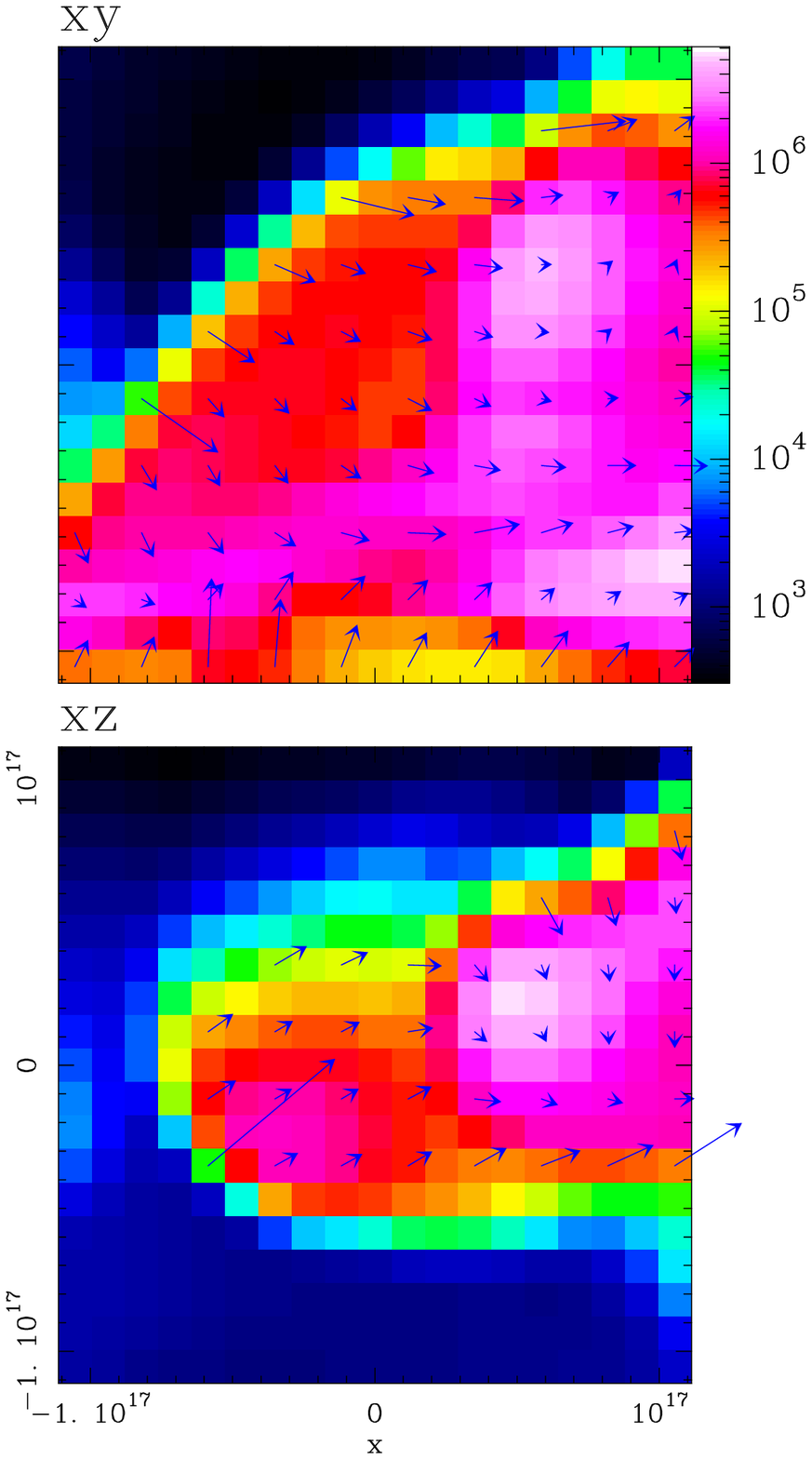}
\caption{Density stratification and flow field in a region
around the centre of mass of the most massive clump at $t=200$~kyr
(see Fig. 6). The top panel shows the flow on a $xy$-cut and the bottom
panel the flow on a $xz$-cut. The origin of the coordinate system
coincides with the centre of mass of the clump (obtained
with a $\rho_c=3\times 10^{-19}$~g~cm$^{-3}$ density cutoff,
which corresponds to a number density of $1.4\times 10^5$~cm$^{-3}$).
The colour scale shows the logarithmic
density distribution (given in cm$^{-3}$ by the bar on the top
right). The arrows show the flow velocities on the planes of the
two cuts for densities  $\rho>10^{-19}$~g~cm$^{-3}$.
An arrow with a length equal to the separation between
successive arrows corresponds to a flow velocity of 0.5~km~s$^{-1}$.}
\end{figure}

\section{Results}
We allowed the model to run from the initial conditions described
in Sect. 2, to a $t=200$~kyr evolutionary time. Figure 1 shows the time
evolution in the mid-plane density stratification without including the
self gravity of the gas. Figure 2 shows the same simulation but adding the
force that arises from self-gravity.

From both Figs.,
it is evident that the ionisation front becomes highly corrugated with
dense condensations at the tip of a number of protruding ``fingers''. 
At $t=100$~kyr, the effect of self-gravity is only to produce 
denser condensations at the tip of the fingers. At $t=200$~kyr, however,
the density structures obtained without (Fig. 1) and with self-gravity
(Fig. 2) are quite different. In the self-gravitating simulation,
a dense, central structure (detached from the ionisation front and
absent in the non-gravitating simulation) is produced.

Given the
important differences found when including self-gravity,
we present an analysis of clump formation only for the
flow obtained from the self-gravitating simulation. Interestingly,
if one repeats the analysis for the non-gravitating simulation,
similar results are found (these results are not shown in the
present paper).

To quantify the number of clumps produced, following
E07 we calculate the number of spatially connected neutral structures with
densities above a specified cutoff density $\rho_c$. In particular,
we choose cutoff values of $\rho_c=3\times$, $10^{-18}$, $10^{-19}$
and $10^{-20}$~g~cm$^{-3}$.

The number of clumps (fragments) obtained
for different density cutoffs, is shown as a function of time in
Fig.~3. To determine the number of clumps, we consider
time intervals of 20~kyr (corresponding to the width of the
bins in the histograms of Fig.~3). We then calculate the number of
clumps in 8 outputs within each of these time intervals, and
then compute an average number of clumps for each time interval.

For the lowest cutoff density ($\rho_c=10^{-20}$), the initial
density distribution has $\sim 80$ clumps, and the number of clumps
first decreases with time, stabilizes at $\sim 5$ for $40<t<170$~kyr,
and then continues to increase a little at $t>170$~kyr
(see Fig.~3). For the intermediate
cutoff density ($\rho_c=10^{-19}$), the initial
distribution has $\sim 500$ clumps and the number of clumps
first decreases and then remains approximately constant as a function
of time (with a value of $\sim 20$).

The initial density distribution has no clumps with densities above
the highest chosen cutoff density, $\rho_c=3\times 10^{-18}$~g~cm$^{-3}$
(see above). Interconnected structures of sufficiently density 
only start to appear at $t\approx 30$~kyr, and their number grows
monotonically with time, stabilizing at a number of $\sim 20$ for
$t>110$~kyr (see Fig. 3). For each of the detected clumps, we first compute the position of the
centre of mass~:
\begin{equation}
 \textbf{R}_{CM}={1\over M}\int_V \rho  \textbf{r}~d^{3}\textbf{x}\,,
\label{r}
\end{equation}
where $V$ is the contiguous volume of the clump and
\begin{equation}
M=\int_V \rho~d^{3}\textbf{x}\,,
\end{equation}
is its mass. We then compute the angular momentum with respect to the centre of
mass of each clump
\begin{equation}
 \textbf{L}=\frac{1}{M}\int \rho~\left(\textbf{r}-\textbf{R}_{CM}\right)
\times \textbf{v}~d^{3}\textbf{x}\,.
\label{l}
\end{equation}
We assume that we observe the computed flow along the $z$-axis (i. e.,
that the $xy$-plane of the computational domain is parallel to the
plane of the sky). The angle
\begin{equation}
\alpha = |\arctan{L_Y/L_X}|\,,
\label{al}
\end{equation}
(with $\vec{L}$ given by equation \ref{l}) then corresponds to the orientation
angle (with respect to the direction of the ionising photon field) of the
angular momentum of the clumps projected onto the plane of the sky. As discussed
in Sect. 1, these directions correspond to the directions in which
bipolar outflows will eventually be ejected when (and if) the clumps
form star+accretion disk systems.

In Fig. 4, we show histograms indicating the fraction of clumps (obtained
for a cutoff density $\rho_c=3\times 10^{-18}$g~cm$^{-3}$)
with different orientations $\alpha$, for the three different elapsed
times ($t=95$, 145, and 200~kyr). For early times, we find that
the $\alpha$ values of the clumps are randomly distributed (between $\sim 
40$ and $180^\circ$). For $t=200$~kyr, $\approx 36$\% of the clumps have
$70^\circ < \alpha < 100^\circ$, and more than $\approx 55\%$ of the clumps have
$60^\circ < \alpha<100^\circ$. From this result, we conclude that the
dense clumps being formed have angular momenta preferentially
aligned in directions perpendicular to the direction of the incident
ionising photon field (which is parallel to the $x$-axis).

The bottom panel of Fig.~5 shows the projected orientation $\alpha$ of the
angular momentum as a function of clump mass for all of the
clumps obtained with the $\rho_c=3\times 10^{-18}$g~cm$^{-3}$
cutoff density, for elapsed times $t=95$ and 200~kyr. 
We see that at $t=95$~kyr most of the clumps (triangles)
have masses $M_c<0.3$M$_\odot$ and angular momenta
with all $\alpha$ orientations.
For $t=200$~kyr (crosses, see bottom panel of Fig. 5), we see
that all of the clumps with $1$M$_\odot<M_c<60$M$_\odot$ have angular
momenta with orientation
angles $70^\circ<\alpha< 120^\circ$. The lower mass clumps (with
$M_c<0.3$M$_\odot$) have angular momenta with
more widely distributed orientations.

We computed the moduli of the specific
angular momenta (i.~e., the momentum per unit mass)
of the clumps.
The values of $L/M$ for $t=95$ and 195~kyr are shown as a function 
of the clump mass $M$ in the top panel of Fig. 5. 
For $t=95$~kyr, we see that the clumps with $M_c<0.3$M$_\odot$ have 
$M/L<6\times 10^{10}$~km$^{2}$s$^{-1}$. The more massive clumps 
(with $M_c>1$M$_\odot$ ) have $M/L>6\times 10^{10}$~km$^{2}$s$^{-1}$. 
For $t=200$~kyr, we see that the $M_c<0.02$M$_\odot$ clumps have 
$M/L<6\times 10^{10}$~km$^{2}$s$^{-1}$,
while the clumps with $M_c>0.02$M$_\odot$
have $M/L>6\times 10^{10}$~km$^{2}$s$^{-1}$.

We now evaluate whether or not these specific angular momenta
have values comparable to those observed in young star systems.
Typical T~Tauri stars have masses $M\approx 2M_\odot$ and 
accretion disks with radii $r_D\approx 50$~AU. The outer Keplerian
orbit of the disk then has a specific angular momentum
$(L/M)_D\approx 5\times 10^{10}$~km$^{2}$s$^{-1}$. This outer orbit
is determined by the material of the highest angular momentum
in the core from which the star+disk system was formed
(see, e.~g., Ulrich 1976).

This value of $(L/M)_D$ is
shown with a horizontal line in the top panel of Fig. 5.
It is clear that many of the clumps formed in our
simulation have specific angular momenta that are
substantially higher than deduced from the
radius of disks around T~Tauri stars. From this, we
conclude that the angular momenta of the clumps generated in our
simulation are substantial.

A relevant question is whether the clumps obtained in our simulations
are resolved well enough for the calculations of angular momenta to
be meaningful. As an example, we consider the clumps found
for the $\rho_c=3\times 10^{-18}$~g~cm$^{-3}$ cutoff
density at time $t=200$~kyr.
The lower mass clumps (see Fig. 5), of $M\approx 0.01$~M$_\odot$,
are resolved with $\sim 5$ grid points. The clumps of
$M\approx 0.1$~M$_\odot$ are resolved with $\sim 50$ grid points.
The clumps with $1<M<100$~M$_\odot$ are resolved with
$\sim 500$ to 5000 grid points. Therefore, for clumps with
$M> 0.1$~M$_\odot$, the resolution of the internal structure
of the clumps (with 50 grid points, corresponding to
$\sim 4$ grid points along each axis) appears to be
appropriate for obtaining a meaningful estimate of the
angular momentum.

From the number of grid points $N_c$ within the clumps, we can
estimate the characteristic radii of the clumps to be $R_c=0.012\,{\rm pc}\,
\times N_c^{1/3}/2$, where 0.012~pc is the resolution of the computational
cells in our numerical simulation. From the values of $N_c$ given
in the previous paragraph, we then see that the clumps obtained
from our simulation have characteristic radii $R_c\approx 0.01$, 0.02,
0.05, and 0.1~pc, for clump masses of $M\approx 0.01$, 0.1, 1, and 100~M$_\odot$,
respectively.

Finally, we study the evolution in the most massive, neutral clump (detected
with $\rho_c=3\times 10^{-18}$~g~cm$^{-3}$). As seen in Fig. 6,
this clump has a mass that
grows monotonically from 0.6~M$_\odot$ at $t=38$~kyr, to 60~M$_\odot$ at
$t=200$~kyr. The orientation angle $\alpha$ (on the plane
of the sky) of its angular momentum stabilizes rapidly at 
$\alpha \approx 90^\circ$ for $t>70$~kyr.

We compute the Jeans mass of this clump to
\begin{equation}
 M_{J}=\frac{1}{6} \pi \bar{\rho} \left( \frac{\pi c_{s}^{2}}{G\bar{\rho}}\right)
 ^{\frac{3}{2}}\,,
\end{equation}
(see, e. g., E07) where $G$ is Newton's constant, $c_s$ is the
sound speed of the neutral medium (see Sect. 2), and $\bar{\rho}=M_c/V^{1/3}$.
We show the ratio
$M_c/M_J$ for the most massive clump as a function of time in the
central panel of Fig. 6. It is clear that this clump is Jeans unstable
for $\sim 160$~kyr, which is a long enough timescale for the formation of
a low mass star.

Figure 7 shows the density and flow velocity distributions in the
$xy$- and $xz$-planes, within a ($2\times 10^{17}$~cm)$^2$ region centred
on the centre of mass of the most massive clump at $t=200$~kyr. In
the two cuts that are shown, we see that the region with
densities higher than the $\rho_c=3\times 10^{-19}$~g~cm$^{-3}$
density cutoff (which corresponds to a number density of
$1.4\times 10^5$~cm$^{-3}$) has a number of density maxima, none
of which coincides with the centre of mass of the structure.
The $xz$-plane (bottom panel) shows the velocity field that gives
rise to the angular momentum of the clump.

\section{Conclusions}
We have presented the results of numerical simulations of a neutral structure
with power-law density perturbations that is photoevaporated by
an incident, plane-parallel, ionising photon field, with and without
the self gravity of the gas. In this interaction,
a number of dense, neutral clumps are produced. Our simulations 
are similar to those presented by Gritschneder
et al. (2009). The main difference is that while they started
their simulations in a medium with turbulent motions, our simulations
begin in a stationary medium with density perturbations. In
our simulations, the velocity field that develops is therefore mainly
the result of the interaction with the ionising photon field. Defining 
clumps as contiguous structures above a cutoff density $\rho_c$,
we compute the statistics of the number of clumps as a function of
elapsed time (for different values of $\rho_c$). We then fix
the cutoff density at $\rho_c=3\times 10^{-18}$~g~cm$^{-3}$, 
to focus on the denser clumps appearing at later elapsed times.

For these clumps, we compute the vector angular momenta, from which
we obtain the direction of the rotation axes (projected on the plane
of the sky) and the specific angular momenta. We find that as a function
of evolutionary time we obtain orientations that are aligned increasingly
perpendicular to the direction of the incident, ionising photon
field.

For the most massive clump, we find that it has a mass that increases across the range
$\approx 0.6$-60~M$_\odot$ (during the $t=38\to 200$~kyr period),
and that the orientation angle $\alpha$ of its angular momentum eventually
stabilizes at $\alpha\approx 90^\circ$ (i. e., the
direction perpendicular to the direction of the incident photon field). We use
an estimate of the Jeans mass of the clump to show that it is
Jeans unstable throughout the $t=38\to 200$~kyr period. This timespan
is long enough for a low mass star to form within the most massive clump.
However, at the resolution of our simulation (with a grid spacing
of $\approx 800$~AU), we naturally do not succeed in form
a star+disk system.

If we analyse our non-gravitating simulation (see Fig. 1), we obtain
qualitatively similar results. Regardless of whether we
consider the self-gravity of the gas or not, we produce clumps with
angular momenta preferentially aligned perpendicular to the direction
of the incident ionising photon field. Even though it is impossible to
provide a full explanation of this alignment, it is possible to provide
a qualitative explanation. During the interaction of the ionising photon
field with a perturbed density structure, a corrugated ionisation front
is produced. This ionisation front pushes a shock into the
neutral gas, producing a sheared velocity field that is preferentially
aligned with the $x$-axis (i. e., with the direction of the ionising
photon field). This sheared velocity field eventually produces vortical
motions that are perpendicular to both the $x$-axis and to the direction of
the shear. This motion is seen in the $xz$-plane velocity field around
the most massive clump in the $t=200$~kyr frame shown in the bottom frame
of Fig.~7.

We have shown that the dense clumps that form
as the result of the photoevaporation of a dense, neutral structure in the ISM
have angular momenta preferentially aligned in a direction
perpendicular to the external ionising photon field. This result provides a natural
explanation of the orientations observed in the \object{HH~555} (Bally \& Reipurth 
2003), \object{HH~666} (Smith, Bally \& Brooks 2004), and \object{HH~333}
(Rosado et al. 1999; Yusef-Zadeh et al. 2005) outflows, which emerge from 
elephant trunks in directions approximately perpendicular to the body of 
the trunks. Future observations of HH flows emerging from externally 
photoionised, neutral structures will show whether or not this kind of 
orientation is a general property of these outflows.

We note again that we have simulated an ionisation front
travelling into an initially steady, neutral medium with density
perturbations. In this way, our simulations follow the dynamics produced
by the propagating ionisation front and associated shock waves, which
result in the production of clumps with angular momenta preferentially
aligned perpendicular to the direction of the ionising photon source.
In the real ISM, a medium with density perturbations also has associated
motions, and an initial vorticity field that will influence the
angular momenta of clumps that might form (e.~g., in the interaction
with an ionisation front). If the initial vorticity field is strong
enough, it will probably hide the effect of the vorticity generated
by the shocks associated with the ionisation front, and the
angular momentum alignment effect described in this paper will
not be present.

An evaluation of whether or not the vorticity generated
by the ionisation front will be hidden by the initial vorticity field
of the cloud (present before the perturbations associated with the
approaching ionisation front) can be completed on the basis of observations
of the rotation of dense clumps in molecular clouds. For example,
Ohashi et al. (1997) observed the kinematics of a number of NH$_3$ cores
in IRAS 04169+2702 and computed their specific angular momenta. They
find that cores with radii in the $0.02\to 0.1$~pc range have specific angular
momenta $L/M\sim (0.3\to 3)\times 10^{11}$~km$^2$s$^{-1}$ (clumps
with larger radii having specific angular momenta up to an order
of magnitude higher for a $\sim 1$~pc clump radius).

In our simulation, the clumps with radii in the $0.02\to 0.1$~pc
range (corresponding to clump masses in the $0.1\to 100$~M$_\odot$ range,
see Sect. 3), have angular momenta $L/M\sim (0.4\to 20)\times
10^{11}$~km$^2$s$^{-1}$ (see Fig. 5). Therefore, our clumps have angular momenta
with values ranging from the lower $L/M$ values of the cores observed
by Ohashi et al. (1997), up to a factor of $\sim 10$ times higher
than the observed values. This result indicates that if the initial
specific vorticity of the structure in the cloud were comparable to
that of IRAS 04169+2702, the passage of an ionisation front
would generate clumps of considerably higher specific vorticity,
and therefore the angular momentum alignment effect described in this
paper would indeed be present (at least for the more massive, higher
angular momentum clumps).

As a final point, we note that in the simulations presented
in this paper we consider only the photoionisation of a neutral structure.
In the case of the interaction of the radiation of an O star with
a molecular cloud, it is unavoidable that the region outside the
ionisation front will be affected by the FUV radiation from the star,
which at least partially photodissociates the initially molecular
material. Gorti \& Hollenbach (2002) computed models of the
photodissociation of dense clumps, and concluded that clumps
with central column densities $<2\times 10^{22}$~cm$^{-2}$
(for an assumed cold-to-dissociated gas sound speed ratio of $\sim 1/3$)
will be rapidly photodissociated, and disappear as local density
enhancements. In our simulations, the clumps that are produced
have central column densities of $\sim (4,9,23,47)\times 10^{22}$~cm$^{-2}$
for clump masses of 0.01, 0.1, 1, and 100~M$_\odot$, respectively (these
central column densities are estimated by multiplying the clump radii
given in Sect. 3 by the cutoff density of $\sim 1.5\times 10^6$~cm$^{-3}$).
Therefore, in all cases the clumps have high enough column densities
to avoid their dissipation by the incident FUV field.

From the results of Gorti \& Hollenbach (2002), we therefore
conclude that the photodissociation caused by the FUV
field will not destroy the clumps produced in our simulations.
However, the early evolution of the flow (in which high density structures
have not yet formed) might indeed be modified by the presence of a
FUV field. It will therefore be interesting to carry out a future
exploration of the formation of clumps within elephant trunks
in the presence of both a photodissociating and a photoionising
photon field.

\begin{acknowledgements}
We acknowledge support from the CONACyT grant 61547. VL acknowledges
the CONACyT scholarship 194595 and Stu~group. We thank an anonymous
referee for helpful suggestions. We thank Malcolm Walmsley for pointing
out that the observations of angular momenta of cores are relevant
for the present work (giving rise to the four last paragraphs of Sect. 4).

\end{acknowledgements}

\label{lastpage}

\begin{thebibliography}{99}

\bibitem[{Bally}{2007}]{bal07} Bally, J., Reipurth, B.,
Davis, C. J. 2007, Protostars and Planets V, eds. B. Reipurth,
D. Jewitt and K. Keil (Univ. of Arizona Press), p. 215-230
\bibitem[{Bally}{2003}]{bal03}
Bally,J., Reipurth,B. 2003, AJ, 126, 893
\bibitem[{Esquivel}{2003}]{esq03}
Esquivel, A., Lazarian, A., Pogosyan, D., Cho, J. 2003, MNRAS, 342, 325
\bibitem[{Esquivel}{2007}]{esq07}
Esquivel, A., Raga, A. C. 2007, MNRAS, 377, 383 (E07)
\bibitem[{Gahm}{2006}]{gah06}
Gahm, G. F., Carlqvist, P., Johansson, L. B.,
Nikoli\'c, S. 2006, A\&A, 454, 201
\bibitem[{Gorti}{2002}]{hol02}
Gorti, U., Hollenbach, D. 2002, ApJ,573, 215
\bibitem[{Gritschneder}{2009}]{grit09}
Gritschneder, M., Naab, T., Walch, S., Burkert, A., Heitsch, F.
2009, ApJ, 694, L26
\bibitem[{Mellema}{2006}]{mel06}
Mellema, G., Arthur, S. J., Henney, W. J. et. al. 2006, ApJ, 647, 397
\bibitem[{Ohashi}{1997}]{oha97}
Ohashi, N., Hayashi, M., Ho, P. T. P., Momose, M.,
Tamura, M., Hirano, N., Sargent, A. 1997, ApJ, 488, 317
\bibitem[{Raga}{2008}]{rag08}
Raga, A. C., Henney, W., Vasconcelos, J., Cerqueira, A., Esquivel, A.,
Rodr\'\i guez-Gonz\'alez, A. 2008, MNRAS, in press
\bibitem[{Reach}{2009}]{rea09}
Reach, W. T., Faied, D., Rho, J.,Boogert, A.,
Tappe, A., Jarrett, T., Morris, P.,
Cambr\'esy, L., Palla, F., Valdettaro, R.
2009, ApJ, 690, 683
\bibitem[{Rosado}{1999}]{ros99}
Rosado, M., Esteban, C., Lefloch, B., Cernicharo, J.,
Garc\'\i a L\'opez, R. J. 1999, AJ, 118, 2962
\bibitem[{Smith}{2004}]{smi04}
Smith, N., Bally, J., Brooks, K. 2003, AJ, 127, 2793 
\bibitem[{Ulrich}{1976}]{ulr76}
Ulrich, R. K. 1976, ApJ, 210, 377
\bibitem[{Yusef-Zadeh}{2005}]{yus05}
Yusef-Zadeh, F., Biretta, J., Wardle, M. 2005, ApJ, 624, 246

\end{thebibliography}
\end{document}